\documentclass[aps,prl,twocolumn,groupedaddress,nofootinbib,showpacs]{revtex4}
\usepackage[dvips]{graphicx}
\newcommand{\BlackHat}{{\sc BlackHat}}
\newcommand{\PYTHIA}{{\sc Pythia}}
\newcommand{\HELACNLO}{{\sc HELAC-NLO}}
\newcommand{\NLOJet}{{\sc NLOJET++}}
\newcommand{\SHERPA}{{\sc SHERPA}}

\newcommand{\AMEGIC}{{\sc AMEGIC++}}
\newcommand{\COMIX}{{\sc COMIX}}

\newcommand{\ntuple}{{$n$-tuple}}

\newif\ifdraft
\draftfalse
\newif\ifpreprint
\preprinttrue

\def\fig#1{fig.~{\ref{#1}}}
\def\Fig#1{Fig.~{\ref{#1}}}

\def\tab#1{table~{\ref{#1}}}

\def\jet{{\rm jet}}
\def\pt{p_T}
\def\root{{\sc root}}

\def\HTpartonic{{\hat H}_T}

\newbox\charbox
\newbox\slabox
\def\s#1{{      
        \setbox\charbox=\hbox{$#1$}
        \setbox\slabox=\hbox{$/$}
        \dimen\charbox=\ht\slabox
        \advance\dimen\charbox by -\dp\slabox
        \advance\dimen\charbox by -\ht\charbox
        \advance\dimen\charbox by \dp\charbox
        \divide\dimen\charbox by 2
        \raise-\dimen\charbox\hbox to \wd\charbox{\hss/\hss}
        \llap{$#1$}
}}

\begin{document}

\title{
\ifpreprint
\hbox{\rm\small
SB/F/402--11$\null\hskip 1.27cm \null$
UCLA/11/TEP/111$\null\hskip 1.27cm \null$
SLAC--PUB--14837$\null\hskip 1.27cm \null$
IPPP/11/82$\null\hskip 1.27cm \null$
CERN--PH--TH/2011/304\break}
\hbox{$\null$\break}
\fi
Four-Jet Production at the Large Hadron Collider
at Next-to-Leading Order in QCD
}

\author{Z.~Bern${}^a$, G.~Diana${}^b$,
L.~J.~Dixon${}^c$, F.~Febres Cordero${}^d$, 
S.~H{\" o}che${}^c$, D.~A.~Kosower${}^b$,
H.~Ita${}^{a,e}$, D.~Ma\^{\i}tre${}^{f,g}$ and K.~Ozeren${}^a$
\\
$\null$
\\
${}^a$Department of Physics and Astronomy, UCLA, Los Angeles, CA
90095-1547, USA \\
${}^b$Institut de Physique Th\'eorique, CEA--Saclay,
          F--91191 Gif-sur-Yvette cedex, France\\
${}^c$SLAC National Accelerator Laboratory, Stanford University,
             Stanford, CA 94309, USA \\
${}^d$Departamento de F\'{\i}sica, Universidad Sim\'on Bol\'{\i}var, 
 Caracas 1080A, Venezuela\\
${}^e$Niels Bohr International Academy and Discovery Center, NBI, DK-2100 Copenhagen, DK\\
${}^f$Department of Physics, University of Durham, Durham DH1 3LE, UK\\
${}^g$PH Department, TH Unit, CERN, CH-1211 Geneva 23, Switzerland\\
}

\begin{abstract}
We present the cross sections for production of up to four jets at the
Large Hadron Collider, at next-to-leading order in the QCD coupling.
We use the \BlackHat{} library in conjunction with \SHERPA{} and a
recently developed algorithm for assembling primitive amplitudes into
color-dressed amplitudes. We adopt the cuts used by ATLAS in their
study of multi-jet events in $pp$ collisions at $\sqrt{s}=7$~TeV.
We include estimates of nonperturbative corrections and compare to
ATLAS data.  We store intermediate results in a framework that
allows the inexpensive computation of additional results for different
choices of scale or parton distributions.

\end{abstract}

\pacs{12.38.-t, 12.38.Bx, 13.87.-a, 14.70.Hp \hspace{1cm}}

\maketitle


Pure-jet events are abundant at the Large Hadron Collider
(LHC), providing a window onto new strongly interacting
physics~\cite{NewPhysics}.  The wealth of data being accumulated
by the LHC experiments motivates comparisons with
precise theoretical predictions from first principles, based
on a perturbative expansion in quantum chromodynamics (QCD) within the
QCD-improved parton model.  The leading order (LO) contribution in the
QCD coupling, $\alpha_s$, does not suffice for quantitatively precise
predictions, which require at least next-to-leading-order (NLO)
accuracy in the QCD coupling.

\begin{figure}[t]
\includegraphics[clip,scale=0.37]{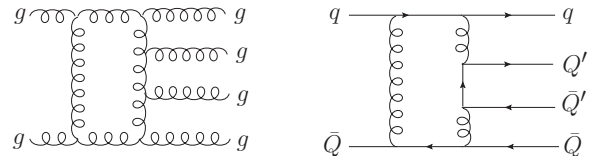}
\caption{Sample diagrams for the six-parton one-loop amplitudes for
 $g g \rightarrow g g g g$ and $q \bar Q \rightarrow  q Q' \bar Q'\bar Q$.}
\label{lcdiagramsFigure}
\end{figure}

The ATLAS~\cite{ATLASJets} and CMS~\cite{CMSJets}
collaborations have recently measured multijet cross sections in $p p$
collisions at 7 TeV. In this Letter, we provide 
NLO QCD predictions for the production
of up to four jets, and compare them to ATLAS data. Our study 
agrees with the earlier two- and three-jet studies performed by
ATLAS collaboration~\cite{ATLASJets} using
\NLOJet~\cite{NLOJet}; the four-jet computation is new.

NLO QCD predictions of jet production at hadron colliders have a
20-year history, going back to the original computations of single-jet
inclusive and two-jet production~\cite{EKS,Jetrad}.  These were
followed by results for three-jet production~\cite{Kilgore,NLOJet}.
A longstanding bottleneck to obtaining NLO predictions for
a larger number of jets at hadron colliders, the evaluation of the
one-loop (virtual) corrections, has been broken by on-shell
methods~\cite{UnitarityMethod,BCFW,OPPEtc}, whose efficiency scales well
as the number of external legs increases.  Recent years have witnessed
calculations with up to five final-state objects~\cite{WZ4j}, among
many other new processes~\cite{PRLW3BH,W3jDistributions,TevZ}.

\begin{table*}
\vskip .4 cm
\begin{tabular}{||c||c|c|c|c|c|c||}
\hline
no. jets & ATLAS &LO &ME+PS  & NLO & NP factor & NLO+NP \\
\hline
\hline
$\geq 2$ & $620 \pm 1.3 {}^{+110}_{-66} \pm 24$ &  $958(1)^{+316}_{-221}$
 & $559 (5)$ &   $1193(3)^{+130}_{-135} $ & $0.95(0.02)$ & 
 $ 1130(19)^{+124}_{-129}$
     \\
\hline
$\geq 3$ & $43 \pm 0.13{}^{+12}_{-6.2}\pm 1.7$  &  $93.4(0.1)^{+50.4}_{-30.3}$  & 
      $39.7 (0.9)$   & $54.5(0.5)^{+2.2}_{-19.9} $  & $0.92(0.04)$
 & $50.2(2.1)^{+2.0}_{-18.3}$
     \\
\hline
$\geq 4 $& $\; 4.3 \pm 0.04{}^{+1.4}_{-0.79} \pm  0.24\;$  
   & $\;9.98(0.01)^{+7.40}_{-3.95}\;$
   & $\;3.97 (0.08)\;$  & $\;5.54(0.12)^{+ 0.08}_{-2.44}\;$ & $\;0.92(0.05)\;$ 
 & $\;5.11 (0.29)^{+0.08}_{ -2.32} \;$ 
 \\
\hline 
\end{tabular} 
\caption{Total cross sections in nb for jet production at the LHC
  at $\sqrt{s} = 7$ TeV,
  using the anti-$k_T$ jet algorithm with $R=0.4$.  We compare ATLAS
  results against LO, ME+PS and NLO theoretical predictions.  The
  penultimate column gives nonperturbative corrections estimated by
  using a string fragmentation model.  In all cases,
  numerical-integration uncertainties are given in parentheses.  The
  scale dependence shown with LO and NLO predictions is given as
  superscripts and subscripts.  The three uncertainties shown with the
  ATLAS data are statistical, jet energy scale, and
  detector unfolding; in addition there is a $\pm 3.4$\%
  luminosity uncertainty.  The jet-energy scale uncertainties are
  asymmetric, so they are given as subscripts and
  superscripts.
\label{CrossSectionAnti-kt-R6Table} 
}
\end{table*}

We illustrate the virtual contributions to four-jet
production in \fig{lcdiagramsFigure}. To evaluate them we have made a
number of significant improvements to the \BlackHat{}
package~\cite{BlackHatI}.  In particular,
assembly of the color-summed cross sections for subprocesses from
primitive amplitudes~\cite{Primitive} 
has been automated~\cite{ItaOzeren},
and the recomputation needed upon detection of numerical
instabilities has been reduced~\cite{HaraldReview}.  The
pure-glue contributions dominate the total cross
section, yet would be the most complex to compute in a traditional
Feynman--diagram approach because of their high tensor rank.
We include all subprocesses and the full color dependence in QCD in
all terms. We treat the five light-flavor quarks as
massless and drop the small (percent-level) effects of top quark loops.

We use \AMEGIC{}~\cite{Amegic}, part of \SHERPA{}~\cite{Sherpa}, to
evaluate the remaining NLO ingredients: the real-emission amplitudes
and the dipole-subtraction terms used to cancel their infrared
divergences~\cite{CS}.
\AMEGIC{} was cross-checked with the \COMIX{} package~\cite{Comix}.
The phase-space integrator exploits QCD antenna
structures~\cite{AntennaIntegrator,GleisbergIntegrator}.

We have carried out extensive checks, including numerical stability,
independence of the phase-space separation parameter 
$\alpha_{\rm dipole}$~\cite{NLOJet}, and cancellation of infrared singularities.
Our results for two- and three-jet production agree with those obtained
by running \NLOJet~\cite{NLOJet} to within 1\%.  (For this comparison 
we used the $k_T$ jet
algorithm~\cite{KTAlgorithm} and CTEQ6M partons~\cite{CTEQ} to match
the default choices in \NLOJet.)  We have compared the virtual matrix
elements for two-, four-, and six-quark processes at selected points
in phase space to \HELACNLO~\cite{HELACNLO}; they agree to 10
digits.  In a supplementary file, we provide reference numerical values
of the virtual matrix elements at a specific phase-space point.

In the fixed-order perturbative expansion of any observable, it is
important to assess whether large logarithms of ratios of physical
scales arise in special kinematic regions. Dijet production, in
particular, suffers from a well-known instability at
NLO~\cite{NLODijetUnsstable}.  If identical cuts on
the transverse momentum $p_T$ of the two jets are used, then
soft-gluon radiation is severely restricted when the leading jet is
just above the minimum $p_T$, while the virtual corrections are
unaffected.  This leads to a large logarithm and a divergence of the
NLO corrections at the minimum $p_T$.  Instead of resumming the
logarithms~\cite{BanfiResum}, we follow ATLAS's approach~\cite{ATLASJets}
of imposing asymmetric cuts, with the minimum $p_T$ of the leading jet
larger than that for additional jets.  Large logarithms are then mitigated
at the price of increased scale dependence for the two-jet prediction:
By $p_T$ conservation, the lowest $p_T$ bins for the first two jets can
only be populated if there is additional real radiation, and the NLO
two-jet prediction effectively becomes LO there.
The production of three or more jets, and in particular the new NLO
prediction for four-jet production, do not suffer from this problem.

In addition to fixed-order parton-level LO and NLO results, we also
present results for a parton-shower calculation matched to fixed-order
LO matrix elements (ME+PS)~\cite{HoechePartonShower}. We obtained the
latter results using a RIVET~\cite{RIVET} analysis within the
\SHERPA{} framework.  We also use \SHERPA{} to estimate
nonperturbative correction factors which we then apply to our NLO
results. These correction factors are obtained by comparing
parton-level results, after showering, to fully hadronized
predictions including a simulation of the underlying event.
We use two different hadronization models: cluster fragmentation as
implemented by \SHERPA{}~\cite{Sherpa} and string fragmentation using
the algorithm in \PYTHIA{} 6.4~\cite{LundFragmentation}.

We consider the inclusive production of up to four jets in $pp$
collisions at a center-of-mass energy $\sqrt{s} = 7$ TeV.  Jets are
defined using the infrared-safe anti-$k_T$ algorithm~\cite{antikT}.
We parallel ATLAS in presenting results for jet-size parameters $R =
0.4$ and $R = 0.6$.  We order the jets in $\pt$.  We implement the
ATLAS cuts from ref.~\cite{ATLASJets}; we require all jets to have
$\pt^\jet > 60$ GeV and the leading jet to have $\pt^{\jet} > 80$ GeV.
Observed jets are also required to have rapidity $|y| < 2.8$.  We use
the MSTW2008 LO and NLO parton distribution functions
(PDFs)~\cite{MSTW2008} at the
respective orders.  We use a five-flavor running
$\alpha_s(\mu)$ and the value of $\alpha_s(M_Z)$ supplied with the
parton distribution functions.

We present our predictions for the LO, ME+PS, and NLO parton-level
inclusive cross sections for two- through four-jet production in
\tab{CrossSectionAnti-kt-R6Table}. 
The strong sensitivity of LO cross sections and distributions to the
variation of the unphysical renormalization scale $\mu_R$ and
factorization scale $\mu_F$ is significantly reduced at NLO.  The wide
range of scales probed in distributions requires us to use an
event-by-event scale characteristic of the kinematics.  We choose
$\mu_R = \mu_F \equiv \mu = \HTpartonic/2$ as our central
scale~\cite{W3jDistributions,TevZ}, where
$\HTpartonic \equiv \sum_i p_T^i$ and the sum runs over all
final-state partons $i$.  We use a
standard procedure to assess scale dependence, varying the central
scale up and down by a factor of two to construct scale-dependence
bands as in ref.~\cite{WZ4j}.  The central scale $\mu = \HTpartonic/2$
is a characteristic measure of the momentum transfers in the event.
It is approximately the jet $\pt$ in the two-jet case, and rises somewhat
in the three- and four-jet cases.  Although it was not tuned in any way,
for three and four jets it happens to lie near the maximum of the
NLO prediction as a function of scale, causing the scale-dependence bands
to be largely to the low side of the central value.  The lowest value
in the band comes from lowering $\mu$ to the lower end of its range,
$\HTpartonic/4$.  (We have not varied
the scale in the ME+PS calculation, as its choice is linked 
to the tuning of various parameters in the parton shower and hadronization
model.  Error sets for these parameters are not available.)  

\begin{table*}
\vskip .4 cm
\begin{tabular}{||c||c|c|c|c|c|c||}
\hline
$p_T$ & ATLAS & LO & ME+PS & NLO & NP factor &  NLO+NP \\
\hline 
60--80 & $ 170 \pm 1.8{}^{+61}_{-33} \pm 12$ & $399 (1)^{+295}_{-157}$ &  $157(4)$  & 
$ 219(6)^{+4}_{-100}$  & $ 0.92(0.06) $ & $202 (14)^{+4}_{-93}$ \\
\hline
80--110 & $24 \pm 0.56{}^{+5}_{-3.8} \pm 2.3$ &  $ 57.6(0.1)^{+42}_{-23} $ &  $23.7(0.7)$ &  
$ 32.6(0.8)^{+0.3}_{-12.9} $ & $0.93(0.05) $ &  $30.3(1.9)^{+0.3}_{-12.0}$\\
\hline
110--160 & $2.6 \pm 0.15{}^{+0.79}_{-0.47} \pm 0.31$ & 
$5.25 (0.01)^{+3.9}_{- 2.1}$&$2.28(0.08)$& $3.3(0.1)^{+0.0}_{-0.9}$  
& $ 0.89(0.06)$ & $2.9(0.2)^{+0.0}_{-0.9}$ \\
\hline
160--210 & $\; 0.15 \pm 0.035{}^{+0.047}_{-0.034} \pm 0.026\;$ & 
 $\; 0.395 (0.001)^{+0.29}_{-0.16}\; $ & 
 $\;0.18 (0.01)\;$ & $\;0.24 (0.01)^{+0.0}_{-0.06}\;$ & $\; 0.93(0.08)$ &
 $\;0.22 (0.02)^{+0.0}_{-0.06}\;$ \\
\hline
\end{tabular}
\caption{The LO, ME+PS and NLO predictions for the distribution
$d \sigma/dp_{T,4}$ [pb/GeV] in the transverse momentum of the 
fourth jet, $p_{T,4}$, for $R = 0.4$, compared to ATLAS data.
The penultimate column gives the nonperturbative correction
factor using the string model.  The final column gives
the NLO prediction including this factor.
\label{ComparisonTable}
}
\end{table*}

In the penultimate column of \tab{CrossSectionAnti-kt-R6Table},
we give the nonperturbative underlying event and
hadronization (NP) correction factor using the \PYTHIA{}-type string
fragmentation model.  The cluster fragmentation model gives
essentially identical results, within our integration uncertainties,
so we do not quote them.  We use this factor as an estimate for the
NP correction to the NLO cross section as well, shown with the
correction in the last column.  (As NLO parton-shower programs are
developed beyond the dijet case~\cite{POWHEGdijets}, 
it will become possible to carry out estimates of
nonperturbative corrections in a manner more compatible with NLO
calculations.)  These nonperturbative corrections are of order 10\% or
less for the production of four or fewer jets. For dijet production, the LO
and NLO theory predictions are not in good agreement with the data; as
discussed above, this is not surprising given the kinematic
constraints as well as the soft-radiation instability.  In contrast,
for the three- and four-jet cases, both the NLO and ME+PS predictions
agree with the data, within the experimental uncertainties, whether or
not we account for the small nonperturbative corrections.

Ratios of cross sections typically reduce both theoretical and
experimental uncertainties. In particular, we have compared the
ratio of four- to three-jet cross sections appearing in 
\tab{CrossSectionAnti-kt-R6Table} to the value obtained by
ATLAS:
\begin{eqnarray}
\hbox{ATLAS:}&&  \hskip -.2 cm 0.098 \pm 0.001{}^{+0.004}_{-0.005} 
                \pm 0.005\,, \nonumber\\
\hbox{ME+PS:} && \hskip -.2 cm   0.100(0.003) \,, \hskip .7 cm 
\hbox{NLO: \hskip .1 cm } 0.102(0.002) \,,   \nonumber
\end{eqnarray}
where the quoted ATLAS uncertainties are respectively statistical, jet
energy scale and detector unfolding~\cite{ATLASJets}. We display only
the statistical integration errors for the theoretical predictions; in
the ratio, the (correlated) scale dependence cancels and is not a
useful estimate of uncertainty.  We have not included the
nonperturbative corrections; they also largely cancel in jet
ratios. We estimate the residual theoretical uncertainty by
comparing ME+PS and NLO results; from here we deduce that the residual
theoretical uncertainty is under 5\%.  This is within our numerical
integration uncertainty and also smaller than the experimental uncertainty.

\begin{figure}[t]
\hbox{$\null$ \hskip -.4 cm \includegraphics[clip,scale=0.55,trim=25 0 0 0]{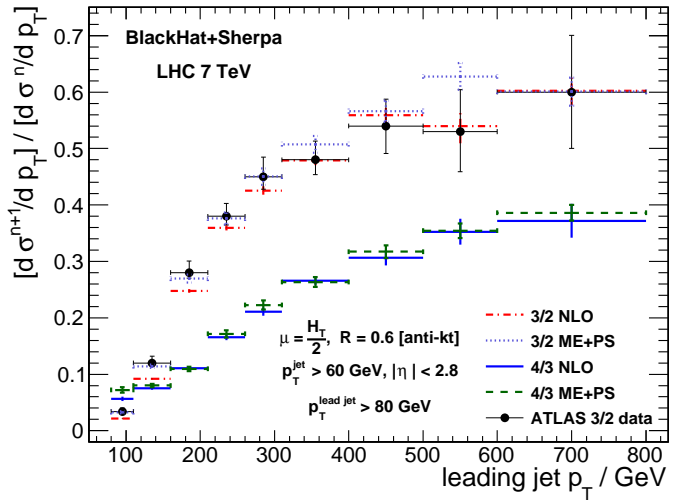}}
\caption{A comparison of the 3/2 and 4/3 jet-production ratios to
  ATLAS data~\cite{ATLASJets} for $R=0.6$.  We show the NLO and ME+PS
  predictions for these ratios. Vertical bars on the theory 
  predictions represent Monte Carlo statistical uncertainties.
\label{RatioFigure}
}
\end{figure}

In \tab{ComparisonTable} we present the LO, ME+PS and NLO $p_T$ distribution
of the fourth-leading jet, comparing to ATLAS data~\cite{ATLASJets}.
The penultimate column gives the nonperturbative 
correction factor, estimated using \SHERPA, as discussed above.  The
final column displays the NLO results including this factor. From this table
we see that both ME+PS and NLO results are in good agreement with the
data, within uncertainties. The estimated nonperturbative corrections
are smaller than current experimental uncertainties.

We also consider the $(n+1)/n$ jet production ratios,
$[d\sigma^{n+1}/dp_T]/[d\sigma^{n}/dp_T]$, as a function of the
leading-jet $\pt$.
\Fig{RatioFigure} displays the 3/2 and 4/3 jet production ratios
for $R=0.6$, comparing the 3/2 ratio with ATLAS data.
For the 3/2 ratio we find very good
agreement between NLO theory and the ATLAS data~\cite{ATLASJets},
except for the first bin, where the denominator is affected by the kinematic
constraint and soft-radiation instability mentioned earlier.
The agreement remains good
even with increasing leading-jet $\pt$, where the ratios grow to $0.6$
and $0.35$ for the 3/2 and 4/3 ratios respectively. 
The ME+PS prediction is also in
very good agreement with data and consistent with NLO, implying that
these processes are under good theoretical control.  It will be
interesting to compare our theoretical predictions for the 4/3 ratio
to future LHC data.  

We have estimated the PDF uncertainty using the 100-element NNPDF 2.1 
error sets; 
the MSTW2008 68\% error sets; and the CT10 90\% CL sets.  
With MSTW2008, we
find one-sigma uncertainties of 1.2\% for two-jet production;
1.6\% for three-jet production; and
2.5\% for four-jet production.  The NNPDF~2.1 and
MSTW08 central values agree to well within these values, and the NNPDF~2.1
one-sigma uncertainties are comparable.
The CT10 PDF uncertainty estimate is about 25\% greater than for MSTW2008.
However, the CT10 central value for three-jet production
is 5.8\% low, outside combined two-sigma errors.
At high $\pt$, the
uncertainty grows somewhat, but remains smaller or comparable to our
numerical-integration errors.

We have studied the dependence of the jet cross sections on the jet
size parameter $R$ for anti-$k_T$.  LO multi-jet cross sections always
decrease with increasing $R$, because whenever two partons are merged the
event is lost.  At NLO, the $R$ dependence is a dynamical question.
We find that the NLO three-jet cross section
increases with $R$ for our usual range of scale variation.
Whether the four-jet cross section increases or decreases with $R$
is sensitive to the choice of scale.

For each event we generate, we record the squared matrix element, the 
momenta of all partons, and the coefficients of various
functions that control the dependence of the final result on the
renormalization and factorization scales, as well as on the PDFs.
We store this
information in \root{}-format \ntuple{} files~\cite{ROOT}.  The
availability of these intermediate results in a standard format makes
it computationally inexpensive to evaluate cross sections and
distributions for different scales and PDF error sets.  They also
offer an easy and reliable way of furnishing our theoretical
predictions to experimental collaborations, while allowing them to
modify cuts or compute additional distributions~\cite{ATLASWconfnote}.

In this study of pure-jet processes, we have imposed cuts
typical of Standard-Model measurements at the LHC.  The same tools
used here can also be used to study backgrounds to new physics
signals, such as those arising from colored resonances or
higher-dimension effective operators. The improved efficiencies
developed in the course of our study should allow us to continue
increasing the number of jets accessible to NLO predictions.


\vskip .3 cm 

We are grateful to Marc-Andre Dufour, Joey Huston, and Brigitte Vachon
for providing us with very helpful information about the ATLAS results
and their comparisons to \NLOJet. We thank the Kavli Institute for
Theoretical Physics, where this work was initiated, for its
hospitality.  This research was supported by the US Department of
Energy under contracts DE--FG03--91ER40662, DE--AC02--76SF00515 and
DE--FC02--94ER40818.  DAK's research is supported by the European
Research Council under Advanced Investigator Grant ERC--AdG--228301.
The work of H.I. and S.H. was partly supported by a grant from the US
LHC Theory Initiative through NSF contract PHY--0705682. This research
used resources of Academic Technology Services at UCLA.

\end{document}